\documentclass[aip,jcp,reprint]{revtex4-2}
 
\usepackage[english]{babel}
\usepackage{braket}
\usepackage{amsthm}
 
\usepackage{mathtools}
\usepackage{siunitx}
\usepackage{accents}
\usepackage{natbib}
\usepackage{xcolor}
\usepackage{amsfonts}

\usepackage{graphicx}% Include figure files
\usepackage{dcolumn}% Align table columns on decimal point
\usepackage{bm}% bold math
\usepackage[utf8]{inputenc}
\usepackage[T1]{fontenc}
\usepackage{mathptmx}
\usepackage{etoolbox}
\usepackage[version=4]{mhchem}
 
\usepackage[normalem]{ulem}

\usepackage[]{xcolor}
\usepackage{soul}
\usepackage{cancel}
\makeatletter
\def\@email#1#2{%
 \endgroup
 \patchcmd{\titleblock@produce}
  {\frontmatter@RRAPformat}
  {\frontmatter@RRAPformat{\produce@RRAP{*#1\href{mailto:#2}{#2}}}\frontmatter@RRAPformat}
  {}{}
}%
\makeatother
\begin{document}

\preprint{AIP/123-QED}

\title[]{Coupled cluster theory for positron binding in anions and polyatomic molecules}
\author{Rosario R. Riso*$^{2}$}
\author{Jan Haakon M. Trabski$^{2}$}
\affiliation{These authors contributed equally to this article}
\author{Federico Rossi}
\affiliation{Department of Chemistry, Norwegian University of Science and Technology, NTNU, 7491 Trondheim, Norway}
\author{Dermot G. Green}
\affiliation{School of Mathematics and Physics, Queen’s University Belfast, Belfast BT7 1NN,
Northern Ireland, United Kingdom}
\author{Henrik Koch*$^{2}$}
%\affiliation{Department of Chemistry, Norwegian University of Science and Technology, NTNU, 7491 Trondheim, Norway}
\email{rosario.r.riso@ntnu.no, henrik.koch@ntnu.no}

\date{\today}

\begin{abstract}
We present the positron coupled cluster singles and doubles (POS-CCSD) method to calculate positron binding energies in molecules. This framework treats electrons and positrons on an equal footing and includes up to simultaneous double-electron–single-positron excitations. We benchmark the approach by computing binding energies for atomic anions and several polar and non-polar polyatomic systems, comparing the results with independent theoretical studies and, where available, experimental data. The fully converged results for H$^{-}$ are in excellent agreement with quantum Monte Carlo and multi-reference configuration interaction results. Quantitative agreement with experiments is not reached in the present study due to the slow convergence of the binding energy with respect to the size of the orbital bases for the electrons and the positron. However, the POS-CCSD results underscore the critical role of electron correlation in the description of electron–positron systems required for a balanced description of these complex systems. In addition, we examine nuclear relaxation effects following positron attachment in LiH.
\end{abstract}

\maketitle

\section{Introduction}
Positrons are the antiparticles of electrons. Their ability to annihilate with atomic and molecular electrons forming characteristic $\gamma$ rays gives them important use as e.g., ultrasensitive probes of defects\cite{RevModPhys.85.1583} and surfaces\cite{HUGENSCHMIDT2016547} in condensed matter and porous materials, in positron and positronium-based medical imaging\cite{PETbook,Moskal2024,Moskal2025:IEEE}, and in astrophysics\cite{RevModPhys.83.1001}. In fundamental physics and chemistry, they are also at the heart of more complicated antimatter, namely positronium (Ps) and antihydrogen, which are now routinely formed and interrogated with the aim to test fundamental symmetries and gravity\cite{cassidy2018, alpha1,alpha2,ATRAP:2012,Amole:2012,Amole:2014,GBAR,Ahmadi:2016,Ahmadi:2017,Malbrunot18,Baker2021,Amsler21,Moskal2021,Adrich2023,Anderson2023}. Moreover, the pioneering development of the buffer-gas positron Surko trap, see e.g., \cite{Murphy:Surko-trap,Danielson:2015}, enabled the routine trapping, accumulation and delivery of energy-tunable positron beams with meV-resolution, paving the way for fundamental studies of atomic and molecular scattering, annihilation and binding.
In particular, measurements of the vibrational-Feshbach resonant annihilation spectra, which arise when a positron attaches whilst simultaneously exciting a vibrational mode of the molecule, have enabled the determination of positron binding energies in $\sim$100 molecules\cite{danielson2010dipole,danielson2012interplay,gribakin2010positron,danielson2025improved,arthur2024positron,danielson2023positron}. This process can lead to intramolecular vibrational redistribution, and can involve non-IR modes, offering a positron-based molecular spectroscopy that is sensitive to both IR and non-IR vibrations \cite{gribakin2010positron}. Proper interpretation of the experiments and advancement of the antimatter-based technologies (traps, beams and PET) requires fundamental understanding of positron interactions with atoms and molecules, and ideally predictive computational capabilities. 

Low-energy positron interactions with atoms and molecules are, however, characterized by strong many-body correlation. They have a pronounced effect, overcoming the positron-nuclear repulsion and leading to modification of scattering cross sections, increases of annihilation rates, and the enhancement and enabling of positron binding
\cite{surko2005low,Amusia:Pos:MBT:He,PhysRevA.52.4541,green2014positron,green2015gamma,hofierka_many-body_2022,Rawlins:2023,cassidy2024many,arthur2024positron},
and positronic bonding\cite{Charry:posbonding2018, Ito:posbonding2020, Bressanini:posbonding:2021a,Bressanini:posbonding:2021b,Bressanini2022-nl,Ito:posbonding:2023,Goli:posbonding2023,Cassidy:2024}.
 They also pose a significant challenge in the description of positron-matter systems and serve as a testbed for development in computational many-body methodologies.  
In molecules, additional complexity arises from the vibrational-Feshbach resonance process that involves coupling of the electronic and vibrational degrees of freedom. This process is described by the Breit-Wigner type model of Gribakin \cite{gribakin2000mechanisms,gribakin2001theory,gribakin2010positron}. However, the \emph{ab initio} calculation of the resonant annihilation spectra remains an open-problem. 
\emph{Ab initio} calculations have primarily addressed the correlation problem, which remains highly challenging in its own right. 
Early theoretical works have been motivated by the prediction of positron binding to neutral atoms \cite{PhysRevA.52.4541} and confirmation by variational calculations \cite{PosLi:1997,Strasburger:1998}, and the experimental progress of the Surko group for molecules. They included methods such as diffusion Monte Carlo\cite{bressanini1998positronium,ito2020first,charry2022correlated}, multireference configuration interaction\cite{saito2003multireference,saito2005multireference} and linearized coupled cluster \cite{harabati2014identification} for atoms and diatomic molecules but their high computational cost makes them unsuitable for polyatomic systems.\cite{strasburger1996quantum,ito2020first,charry2022correlated}
Most approaches focused mainly on polar molecules for which positron binding can occur even at the static level of interaction (positrons bind in the static approximation to a point dipole of $>1.625$ D\cite{Crawford67}). 
Despite extensive measurements ($\sim$100 molecules), until 2022 direct comparison between experiment and \emph{ab initio} theory was only available for 6 molecules, with severely deficient theoretical accuracy (at best 25\% error). The development of the diagrammatic many-body theory approach by Green and co-workers \cite{hofierka_many-body_2022} provided significantly more accurate \emph{ab initio} calculations of positron binding energies in molecules (typically within $\sim$10\% of the experimental values) by accounting for the dominant correlations terms. The approach was also extended to positron scattering and annihilation in H$_2$, N$_2$ and CH$_4$\cite{Rawlins:2023} and annihilation $\gamma$-ray spectra \cite{gregg2025many}. Despite its success, the current implementation of the method accounts for pure electron correlation only via the electron-hole polarisation propagator in the positron-molecule self energy, neglecting pure electron-correlation corrections to the Hartree-Fock wave function of the bare target molecule that high precision calculations should account for. 

An alternative approach that treats both interactions on an equal footing, is the coupled cluster (CC) theory. Coupled-cluster methods are a cornerstone in modern quantum chemistry, renowned for their accurate treatment of electron correlation.\cite{bartlett2007coupled,crawford2007introduction} Their systematic hierarchy allows balancing cost and accuracy, providing highly reliable predictions for ground and excited state energies of molecules at a polynomial cost.\cite{bruno2022can,koch_calculation_1994,koch1990coupled,epifanovsky2015spin,bartlett2007coupled,crawford2007introduction,ruud2002optical,pedersen1997coupled,pedersen2004origin} Moreover, a particular benefit of having a coupled cluster approach for positron-molecule interactions lies in the fact that the method has been successfully extended to strongly correlated light–matter and electron–nuclear systems \cite{haugland_coupled_2020,deprince2021cavity,pavovsevic2023computational}, and may thus enable similar developments in positron physics and chemistry. 
\begin{figure*}
    \centering
    \includegraphics[width=0.85\textwidth]{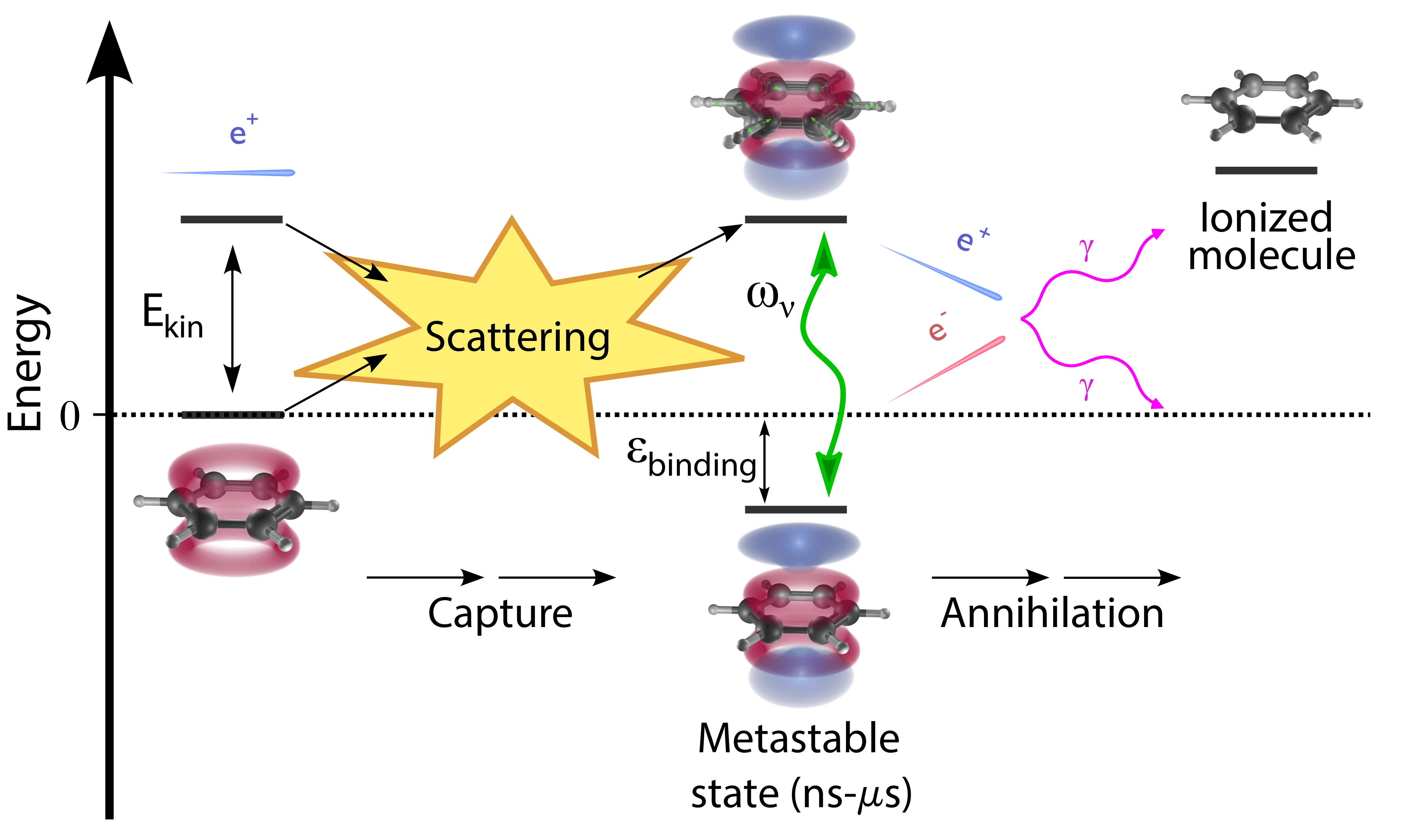}
    \caption{Pictorial representation of the electron-positron capture process. Because of the electron polarization, a bonded meta-stable state (energy minimum) is observed. The dissociation energy is referred to as $\varepsilon_{b}$. 
    Vibrational-Feshbach resonant attachment occurs when the kinetic energy of the incoming positron, $\textrm{E}_{\textrm{kin}}$, plus the binding energy matches a vibrational excitation of the molecule, $\omega_{\nu}$.}
    \label{fig:energetics} 
\end{figure*}

Here, we present a positron-based coupled cluster singles and doubles (POS-CCSD) approach to positron binding in anions and polyatomic molecules, which provides a non-perturbative treatment of both electron–positron and electron–electron correlation. Unlike earlier CC schemes for positron-atom binding, for example the linearized coupled cluster approach by Harabati and co-workers \cite{dzuba2012relativistic,harabati2014identification}, the present method retains the full CC expansion, including complete double excitations in both the electron–electron and electron–positron operators.
Our POS-CCSD approach follows the prescription of the multicomponent CCSD by Chakraborty and co-workers\cite{ellis2016development} employs working equations analogous to the nuclear–electronic orbital coupled cluster singles and doubles (NEO-CCSD) method by Hammes-Schiffer and co-workers\cite{fetherolf2025nuclear,pavovsevic2019multicomponent} and Brorsen and co-workers\cite{fowler2022t}. However, despite these formal similarities, applications of CCSD-level multicomponent methodologies to polyatomic systems have, to the best of our knowledge, not been reported previously. The present work therefore extends these approaches to a new class of chemically relevant systems and discusses their applicability to polyatomic molecules.\\
This paper is organized as follows. Section~\ref{sec:Theory} introduces the positron-electron Hamiltonian and the POS-CCSD methodology. The method is first applied to the atomic ions H$^-$ and F$^-$. Although there is no experimental data for these systems, accurate Monte Carlo and multireference configuration interaction results provide benchmarks, to which our calculations are found to be in good agreement with. Finally, we present binding energy predictions for polar and non-polar polyatomic molecules and compare them with experimental results. This is followed by a discussion of positron-aided nuclear relaxation effects and possible repercussions on the comparison between theoretical and experimental numbers. The concluding section provides final remarks and perspectives.

\section{Theory}\label{sec:Theory}
In the Born-Oppenheimer approximation, the Hamiltonian for the positron + $N$-electron molecule system, $H$, can be written as the sum of three different contributions\cite{hofierka_many-body_2022,pavovsevic2019multicomponent}
\begin{equation}
H=H_{e}+H_{p}+H_{pe},    
\end{equation}
where $H_{e}$ is the electronic Hamiltonian, $H_{p}$ is the positron Hamiltonian and $H_{pe}$ is the interaction term between positrons and electrons. In second quantization,\cite{helgaker2013molecular} these operators are rewritten as
\begin{equation}
\begin{split}
H_{e} = & \sum_{pq}h_{pq}E_{pq}+\frac{1}{2}\sum_{pqrs}g_{pqrs}(E_{pq}E_{rs}-\delta_{rq}E_{ps})\\
H_{p} = & \sum_{PQ}h_{PQ}E_{PQ}+\frac{1}{2}\sum_{PQRS}g_{PQRS}(E_{PQ}E_{RS}-\delta_{RQ}E_{PS})\\
H_{pe} = & \sum_{PQpq}g_{PQpq}E_{PQ}E_{pq},   
\label{eq:full} 
\end{split}
\end{equation}
where lower case letters denote electron orbitals while capital letters denote positron orbitals. The singlet operators for electrons $E_{pq}$ and positron $E_{PQ}$ are defined as
\begin{equation}
\begin{split}
E_{pq} = \sum_{\sigma}a^{\dagger}_{p\sigma}a_{q\sigma} \hspace{1cm}
E_{PQ} = \sum_{\sigma}c^{\dagger}_{P\sigma}c_{Q\sigma}.\label{eq:Second_quantized}
\end{split}
\end{equation}
In Eq.(\ref{eq:Second_quantized}), $a^{\dagger}_{p\sigma}$ creates and $a_{p\sigma}$ annihilates an electron in the spatial orbital $p$ with spin $\sigma$ while $c^{\dagger}_{P\sigma}$ and $c_{P\sigma}$ is similar for positrons. The standard one- and two-body integrals $h$ and $g$\cite{helgaker2013molecular} have been extended to include positron interactions
\begin{equation}
\begin{split}
h_{pq} =& \int \phi^{*}_{p}(r)\left(\frac{\nabla^{2}}{2}-\frac{Z_{\alpha}}{\left|r-R_{\alpha}\right|}\right)\phi_{q}(r)d^{3}r,  \\
h_{PQ} =& \int \phi^{*}_{P}(r)\left(\frac{\nabla^{2}}{2}+\frac{Z_{\alpha}}{\left|r-R_{\alpha}\right|}\right)\phi_{Q}(r)d^{3}r,  \\
g_{PQpq} =& -\int \frac{\phi^{*}_{P}(r)\phi_{Q}(r)\phi^{*}_{p}(r^{\prime})\phi_{q}(r^{\prime})}{\left|r-r^{\prime}\right|}d^{3}r \;d^{3}r^{\prime},  \\
g_{pqrs} =& \int \frac{\phi^{*}_{p}(r)\phi_{q}(r)\phi^{*}_{r}(r^{\prime})\phi_{s}(r^{\prime})}{\left|r-r^{\prime}\right|}d^{3}r \;d^{3}r^{\prime},  \\
g_{PQRS} =& \int \frac{\phi^{*}_{P}(r)\phi_{Q}(r)\phi^{*}_{R}(r^{\prime})\phi_{S}(r^{\prime})}{\left|r-r^{\prime}\right|}d^{3}r \;d^{3}r^{\prime}.
\end{split}
\end{equation}
We only consider cases in which one positron is captured per molecule because the coulomb repulsion with an already captured positron makes a two positron capture extremely unlikely. Therefore the positron Hamiltonian in Eq.(\ref{eq:full}) can be simplified to:
\begin{equation}
H_{p}=\sum_{PQ}h_{PQ}E_{PQ}.    
\end{equation}
Throughout this paper the positron-molecule binding energy $\varepsilon_{b}$
%for the positron 
will be computed as the difference between the positron-molecule complex ground state energy and the energy of the ground state molecule (in the absence of the positron),
where we point out that both should be, for accuracy, computed at the same level of theory and with the same basis set to avoid superposition errors.
\subsection{Positron Hartree-Fock}

\begin{figure}
    \includegraphics[width=0.5\textwidth]{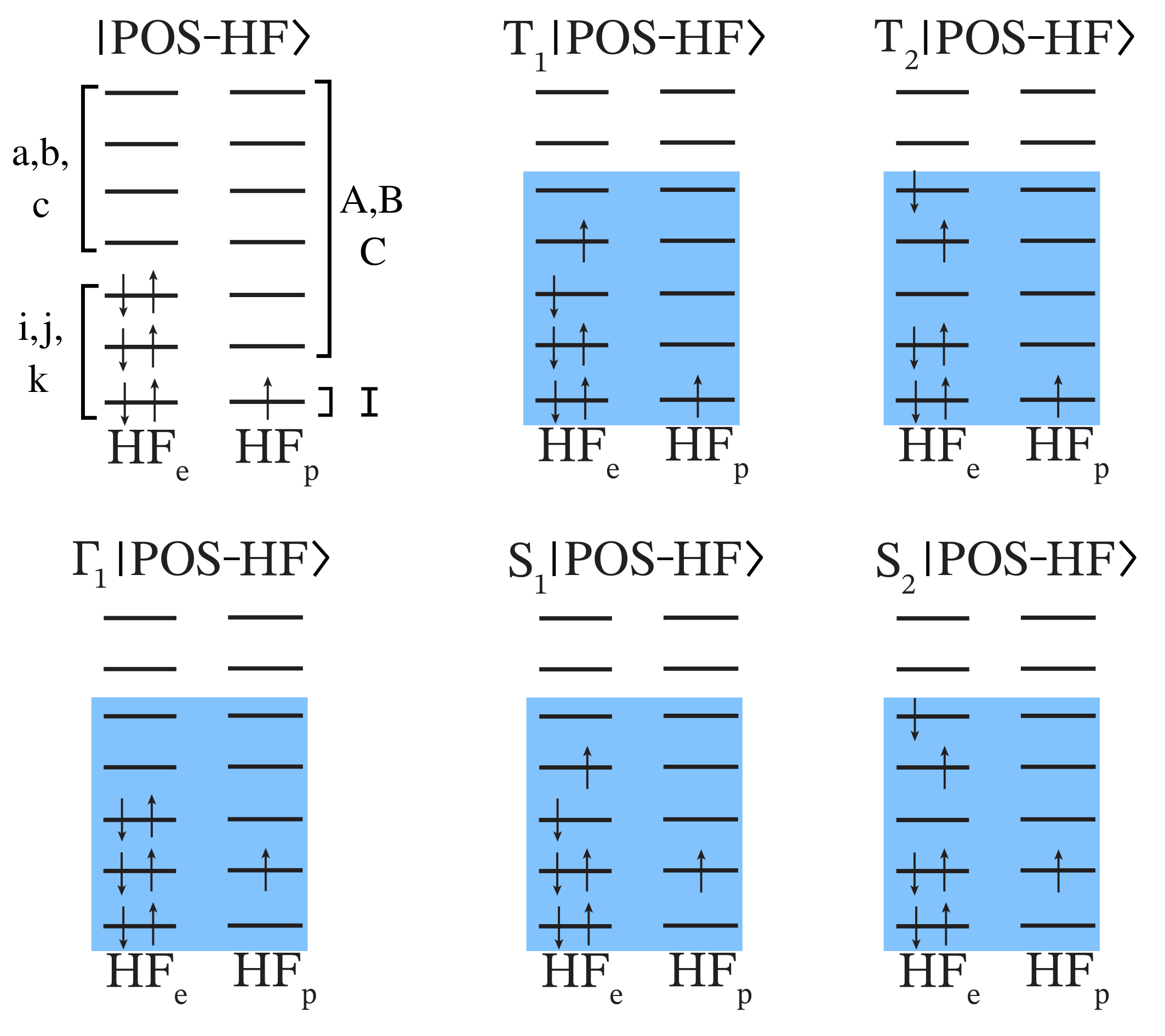}
    \caption{Pictorial representation of the positron Hartree-Fock wave function and the effect of the excitation operators in the cluster on the positron HF wave function. The electronic part $\ket{\textrm{POS-HF}}$ is a Slater determinant where $\alpha$ and $\beta$ electrons occupy the first $N_{e}/2$ orbitals, with $N_{e}$ the number of electrons in the system. The single positron occupies the lowest energy orbital I. The effect of the excitation operators T$_{1}$ and T$_{2}$ is to move one or two electrons from the occupied orbitals to the virtual orbitals,  $\Gamma$ excites the single positron, while $S_1$ and $S_2$ generate simultaneous electron-positron excitations corresponding to single-electron-single-positron and double-electron-single-positron excitations, respectively.
    An active space restriction in the cluster operator can be obtained by selecting a restricted set of orbitals to which the particles can be excited. For example, in this figure only the orbitals in the light blue panels would be 
    %featured 
    included
    in the cluster indices used in the POS-CC calculation}.
    \label{fig:CC_and_states}
\end{figure}
The Hartree-Fock (HF) method is the foundational starting point for most electronic-structure theories, providing the basic mean-field description upon which modern correlation methods are built.\cite{bartlett1994applications} The main idea of the approach is that each particle occupies a single molecular orbital determined by the mean-field potential of the other particles in the system, see Fig.\ref{fig:CC_and_states}. In positron–molecule calculations, two variants of the Hartree–Fock approximation are commonly used: the “relaxed-target” form\cite{swann_calculations_2018}, used in this work, in which the electronic orbitals respond to the presence of the positron, and the “frozen-target” form, in which the electronic HF state is kept fixed and only the positron orbital is optimized.\cite{Amusia:CSAP,amusiabook,Amusia:Pos:MBT:He,PhysScripta.46.248,dzuba_mbt_noblegas,swann_calculations_2018,Gribakin:2004,green2014positron,hofierka_many-body_2022} In the relaxed-target case for a closed-shell $N$-electron molecule positron system, the positron Hartree-Fock (POS-HF) wave function is written as the direct product of an electronic Slater determinant and a positron wave function
\begin{equation}
    \ket{\text{POS-HF}} = \exp(-\kappa)\prod_{i}^n a^{\dagger}_{i\alpha}a^{\dagger}_{i\beta}\otimes  \exp(-\textrm{K})c^{\dagger}_{I \sigma}\ket{\text{vac}},    
\end{equation}
where $\ket{\textrm{vac}}$ is the electron and positron vacuum 
and the optimal system (N electrons plus positron) wave function is obtained by minimizing the POS-HF energy
\begin{equation}
E_{\text{POS-HF}} = \sum_{i}2h_{ii}+\sum_{ij}(2g_{iijj}-g_{ijji})+h_{II}+\sum_{j}2g_{IIjj}\label{eq:p_hf_energy}
\end{equation}
with respect with respect to the real antisymmetric operators $\kappa$ and $\textrm{K}$ 
\begin{equation}
\begin{split}
\kappa =& \sum_{ai}\kappa_{ai}(E_{ai}-E_{ia}) \hspace{1cm} \kappa_{ai} \in \mathbb{R} \\
\textrm{K}=&\sum_{A}\textrm{K}_{AI}(E_{AI}-E_{IA})\hspace{0.6cm} \textrm{K}_{AI} \in \mathbb{R} .\label{eq:prime}
\end{split}
\end{equation}
In agreement with the standard notation, in the remainder of this paper we use the letters $i,j,k...$ to denote occupied electronic orbitals and $a,b,c...$ to denote virtual orbitals. The notation for the positron orbitals will follow the same rules, albeit with capital letters. 
We note that in Eq~(\ref{eq:prime}) only a summation over $A$ is considered because there is only one positron in the system.
The derivative of the energy with respect to $\kappa_{ai}$ and $\textrm{K}_{AI}$ defines the off-diagonal elements of the electron and positron Fock operators
\begin{equation}
\begin{split}
F_{pq} & = h_{pq}+\sum_{i}(2g_{pqii}-g_{piiq})+g_{IIpq} \\
F_{PQ} & = h_{PQ}+2\sum_{i}g_{PQii}, \label{eq:Focks}
\end{split}
\end{equation}
which are used to minimize the energy in Eq.(\ref{eq:p_hf_energy}) via the Roothan Hall equations.
We note that the electron and positron problems are coupled to each other and therefore need to be solved self-consistently. Furthermore, diagonalization of the two Fock matrices in Eq.(\ref{eq:Focks}) provides the canonical set of electron and positron orbitals, to be used in the subsequent coupled cluster calculations. Note that the alternative frozen-target approach  corresponds to optimizing the electron Slater determinant independently from the positron, i.e. removing $g_{IIpq}$ from the Fock matrix in the first line of Eq.(\ref{eq:Focks}).\cite{hofierka_many-body_2022,swann_calculations_2018,amusia2013atomic} 
\section{Positron coupled cluster}
In the coupled cluster approach, the $N$-electrons positron wave function is expressed as the exponential of an excitation operator acting on a reference state, typically a POS-HF determinant. This exponential form ensures the size-extensivity of the method, a critical property for accurate modeling of large systems.
Within our positron coupled cluster (POS-CC) method, the $N$-electron-positron wave function is written as
\begin{equation}
    \ket{\text{POS-CC}} = \exp(T) \ket{\text{POS-HF}} = \exp(T)\ket{\text{HF}_{\text{e}},\text{HF}_{\text{p}}}, \label{eq:p_CCSD}
\end{equation}
where the cluster operator $T$ includes electron, positron, and electron-positron excitation operators. In this work, the cluster operator includes single and double electronic excitations and up to single excitations in the positron (POS-CCSD), see Fig.~\ref{fig:CC_and_states}. Specifically, the $T$ operator is given by 
\begin{align}
T = & T_{1}+T_{2}+S_{1}+S_{2}+\Gamma = \sum_{\mu}t_{\mu}\tau_{\mu} \label{eq:cluster}
\end{align}
where $\mu$ are the excited states included in the cluster definition and 
\begin{align} 
T_{1} = & \sum_{ai}t^{a}_{i}E_{ai} \hspace{1cm} 
T_{2} = \frac{1}{2}\sum_{aibj}t^{ab}_{ij}E_{ai}E_{bj} \\
S_{1} = & \sum_{Aai}s^{Aa}_{Ii}E_{ai}E_{AI} \hspace{0.34cm} 
S_{2} = \frac{1}{2}\sum_{Aaibj}s^{Aab}_{Iij}E_{ai}E_{bj}E_{AI} \\ 
\Gamma =& \sum_{A}\gamma^{A}_{I}E_{AI} \hspace{1cm} t^{a}_{i},t^{ab}_{ij},s^{Aa}_{Ii},s^{Aab}_{Iij},\gamma^{A}_{I}\in \mathbb{R}.  
\end{align}
The amplitudes 
$t^{a}_{i},t^{ab}_{ij},s^{Aa}_{Ii},s^{Aab}_{Iij}$ and $\gamma^{A}_{I}$ are determined requiring that 
\begin{align}
    \Omega_{\mu} =& \bra{\mu} \bar{H} \ket{\textrm{POS-HF}} = 0 \\
    \ket{\mu}=&\tau_{\mu}\ket{\textrm{POS-HF}},\label{eq:omega}
\end{align}
where the similarity transformed Hamiltonian
\begin{equation}
\bar{H}= \exp(-T)H\exp(T),   \label{eq:Similarity_transformed} 
\end{equation} 
has been introduced.
The detailed expression of Eq.(\ref{eq:omega}) is reported in the Supplementary Material. The POS-CCSD energy is obtained as the expectation values of Eq.(\ref{eq:Similarity_transformed}) with the POS-HF state
\begin{equation}
\begin{split}
    E =& \bra{\textrm{POS-HF}} \bar{H} \ket{\textrm{POS-HF}}\\
      =&E_{\textrm{POS-HF}} + \sum_{aibj}(t^{ab}_{ij}+t^{a}_{i}t^{b}_{j})(2g_{aibj}-g_{ajbi}) \\
      +& \sum_{A}h_{IA}\gamma^{A}_{I}+\sum_{Aai}2g_{IAia}(t^{a}_{i}\gamma^{A}_{I}+s^{Aa}_{Ii}).     
\end{split}
\end{equation}
Note that the energy does not depend on the $S_{2}$ amplitudes explicitly, but implicitly 
as they enter the $\Omega$ equations determining $S_{1}, \Gamma,T_{1}$ and $T_{2}$. A detailed discussion of the main differences between POS-CCSD and the other relevant methods presented in the literature is reported in the Supplementary Materials.

\section{Results}
In this section, we report the positron binding energies for a selection of negative ions and molecules as computed using POS-HF and POS-CCSD. Both methods have been implemented in a development version of the eT program.\cite{folkestad20201,folkestad20252} An independent Julia implementation of POS-CCSD obtained using the automatic code generator SpinAdaptedSecondQuantization.jl\cite{lexander2025spinadaptedsecondquantization} was also used to help validate the eT implementation. 
\subsection{Single-atom  anion binding energies}\label{sub:sec:ion}
\begin{table*}[]
    \centering
\caption{Binding energies in eV for H$^{-}$ at the POS-HF and POS-CCSD level computed using standard aug-cc-pVnZ and d-aug-cc-pVnZ basis sets. To showcase the importance of the S$_{2}$ excitation operator we also report the binding energy without the its inclusion in the cluster operator.\footnote{Reference value, from quantum Monte Carlo calculation is $\varepsilon_b$=7.11 eV\cite{ito2020first}.}}
\begin{tabular}{c c c c c c c c c c c c c c}
\hline
& \multicolumn{4}{c}{POS-HF}
& \multicolumn{4}{c}{POS-CCSD without S$_2$}
& \multicolumn{4}{c}{POS-CCSD}
& \\
\cline{1-13}
n
& \multicolumn{1}{c}{aug-cc-pVnZ} & & \multicolumn{1}{c}{d-aug-cc-pVnZ} & 
& \multicolumn{1}{c}{aug-cc-pVnZ} & & \multicolumn{1}{c}{d-aug-cc-pVnZ} & 
& \multicolumn{1}{c}{aug-cc-pVnZ} & & \multicolumn{1}{c}{d-aug-cc-pVnZ} & \\
\hline
2 & 4.817\;\textrm{eV} & & 4.823\;\textrm{eV} 
  & & 5.673\;\textrm{eV} & & 5.736\;\textrm{eV} 
  & & 5.728\;\textrm{eV} & & 5.812\;\textrm{eV} \\
3 & 4.840\;\textrm{eV} & & 4.846\;\textrm{eV} 
  & & 6.121\;\textrm{eV} & & 6.218\;\textrm{eV} 
  & & 6.231\;\textrm{eV} & & 6.357\;\textrm{eV} \\
4 & 4.853\;\textrm{eV} & & 4.858\;\textrm{eV} 
  & & 6.336\;\textrm{eV} & & 6.447\;\textrm{eV} 
  & & 6.471\;\textrm{eV} & & 6.616\;\textrm{eV} \\
5 & 4.870\;\textrm{eV} & & 4.866\;\textrm{eV} 
  & & 6.458\;\textrm{eV} & & 6.569\;\textrm{eV} 
  & & 6.634\;\textrm{eV} & & 6.756\;\textrm{eV} \\
6 & 4.868\;\textrm{eV} & & 4.869\;\textrm{eV} 
  & & 6.532\;\textrm{eV} & & 6.638\;\textrm{eV} 
  & & 6.697\;\textrm{eV} & & 6.836\;\textrm{eV} \\
\hline
\end{tabular}
\label{tab:H_ion_single}
\end{table*}
\begin{table*}[]
\centering
\caption{Binding energies in eV for F$^{-}$ at the POS-HF and POS-CCSD level computed using standard quantum chemistry basis sets. Similarly to what was observed for H$^{-}$, correlation plays a very significant role in determining the binding energy. The property is not well converged with respect to the basis set. Also for F$^{-}$ the inclusion of the $S_{2}$ operator improves the accuracy of $\varepsilon_{b}$\footnote{Reference value, from multireference CI calculation is $\varepsilon_{b}=6.23$ eV\cite{saito2005multireference}}.}
\begin{tabular}{c c c c c c c c c}
\hline
& \multicolumn{2}{c}{POS-HF} & & 
  \multicolumn{2}{c}{POS-CCSD without S$_2$} & & 
  \multicolumn{2}{c}{POS-CCSD} \\
\cline{2-3} \cline{5-6} \cline{8-9}
n
& aug-cc-pVnZ & d-aug-cc-pVnZ 
& & aug-cc-pVnZ & d-aug-cc-pVnZ 
& & aug-cc-pVnZ & d-aug-cc-pVnZ \\
\hline
2 & 4.024\;\textrm{eV} & 4.975\;\textrm{eV} 
  & & 4.623\;\textrm{eV} & 5.509\;\textrm{eV} 
  & & 4.717\;\textrm{eV} & 5.647\;\textrm{eV} \\
3 & 4.309\;\textrm{eV} & 5.004\;\textrm{eV} 
  & & 5.184\;\textrm{eV} & 5.767\;\textrm{eV} 
  & & 5.329\;\textrm{eV} & 5.948\;\textrm{eV} \\
4 & 4.466\;\textrm{eV} & 5.013\;\textrm{eV} 
  & & 5.470\;\textrm{eV} & 5.891\;\textrm{eV} 
  & & 5.652\;\textrm{eV} & 6.096\;\textrm{eV} \\
5 & 4.593\;\textrm{eV} & 5.016\;\textrm{eV} 
  & & 5.648\;\textrm{eV} & 5.950\;\textrm{eV} 
  & & 5.852\;\textrm{eV} & 6.160\;\textrm{eV} \\
6 & 4.730\;\textrm{eV} & --- 
  & & 5.782\;\textrm{eV} & --- 
  & & 5.997\;\textrm{eV} & --- \\
\hline
\end{tabular}
\label{tab:F_ion_single}
\end{table*}
Because of their net negative charge, negative ions host strongly bound positron states that have been described with high accuracy and reproducibility across multiple methodologies like Diffusion Monte Carlo,\cite{ito2020first,charry2022correlated,bressanini1998positronium} multi-reference CI \cite{saito2005multireference} and many-body methods.\cite{Ludlow10,romero2014calculation,hofierka_gaussian-basis_2023-1} While the ion-positron states have not yet been realised experimentally, calculating their binding energies offers a way to benchmark computational approaches and to gain insight into electron–positron correlation. In Tables \ref{tab:H_ion_single} and \ref{tab:F_ion_single}, we report the POS-HF and POS-CCSD binding energies for both H$^{-}$ and F$^{-}$ computed using standard quantum chemistry basis sets. We highlight that for the H$^{-}$ case POS-CCSD is exact within a given basis set, i.e it captures all the correlation in the two electron plus positron system. 
The reference theoretical values are $\varepsilon_{b}=7.11$ eV\cite{ito2020first} for H$^{-}$ and $\varepsilon_{b}=6.23$ eV\cite{saito2005multireference} for F$^{-}$ from QMC and multireference CI calculations, respectively. As expected, for both ions the POS-HF binding energies are significantly underestimated. 
In the POS-CCSD case even large basis set calculations with high angular momentum functions fail to exactly describe the $\textrm{H}^{-}$ binding, with the error decreasing to 0.28 eV for the d-aug-cc-pV6Z basis. Better results are obtained for F$^{-}$, where the d-aug-cc-pV5Z basis set already leads to an error of 0.07 eV.

To gauge the importance of the double electron-positron excitation operator in the description of positron binding, we report in Tabs. \ref{tab:H_ion_single} and \ref{tab:F_ion_single} the binding energies computed without including $S_{2}$ in the cluster operator. We note here that despite not being directly featured in the energy expression, the $S_{2}$ amplitude implicitly affects the other amplitudes, and its inclusion improves the predicted binding energies for all the considered basis sets. The effect of $S_{2}$ increases with the basis set size, reaching around 200 meV in the aug-cc-pV6Z basis. While a 200 meV error is lower than 5$\%$ of the reference binding energy, we emphasize that neglecting the $S_{2}$ contribution doubles the theoretical error. In the remainder of the paper all calculations will therefore include $S_{2}$ in the cluster operator. \\

\begin{table}[]
\centering
\caption{Binding energies in eV for H$^{-}$ at the POS-CCSD level computed using optimized aug-cc-pVnZ basis sets. We notice that the results improve very significantly for small n (i.e. low angular momenta) and more mildly for larger n. Nevertheless, significant improvements are observed in the predicted binding energies.}
\begin{tabular}{c l c l c}
\hline 
n && {aug-cc-pVnZ} && {Opt-aug-cc-pVnZ} \\
\hline
2 && 5.728\;\textrm{eV} && 5.998\;\textrm{eV} \\
3 && 6.231\;\textrm{eV} && 6.507\;\textrm{eV} \\
4 && 6.471\;\textrm{eV} && 6.732\;\textrm{eV} \\
5 && 6.634\;\textrm{eV} && 6.839\;\textrm{eV}\\
6 && 6.697\;\textrm{eV} && 6.842\;\textrm{eV} \\
\hline 
\end{tabular}
\label{tab:optimized}
\end{table}
Since POS-CCSD is formally exact for two electrons and one positron, the discrepancy between the POS-CCSD binding energies and the theoretical reference must be connected to an inadequacy of the basis set. 
The positron is indeed highly diffuse around the ions and standard electronic basis sets are not able to 
properly describe the full spatial extent of the positron wavefunction.
To further emphasize this point, we report in Tab. \ref{tab:optimized} how the binding energies change if the aug-cc-pVnZ Gaussian exponents are optimized for the positron-$\textrm{H}^{-}$ using the software package BasisOpt.\cite{shaw2023basisopt} 
The predicted binding energies improve when optimizing the exponents, in particular for low angular momentum basis sets(n). Nonetheless we observe that the improvement in the description of $\varepsilon_{b}$ is saturated quickly and the effect of optimization is not enough to reconcile the POS-CCSD binding energy with the theoretical reference. \\
\begin{table}[]
    \centering
     \caption{Binding energies (in eV) at different active space dimensions for F$^{-}$ and H$^{-}$ using additional ghost atoms. As the active space dimension increases, the positron becomes more tightly bound. However, very large active spaces are needed for saturated results. The full space number of orbitals for H$^{-}$ is 1300 and 1800 for F$^{-}$.}
    \label{tab:Ghost_atoms}
    \begin{tabular}{c c c }
    \hline 
     Active space dimension & H$^{-}$ & F$^{-}$ \\
    \hline
    300 & 6.804\;\textrm{eV} & 6.099\;\textrm{eV} \\
    500 & 6.995\;\textrm{eV} & 6.235\;\textrm{eV} \\
    700 & 7.051\;\textrm{eV} & --- \\
    Ref.\cite{saito2003multireference,saito2005multireference,bressanini1998positronium,ito2020first} & 7.110\;\textrm{eV} & 6.230\;\textrm{eV} \\ 
    \hline 
\end{tabular}
\end{table}

A widely used solution to tackle this problem is to combine multiple basis sets to account for all the main physical effects in the positron-molecule system. Specifically, the correlation consistent augmented basis sets are optimized to accurately model the electronic wave function while even tempered gaussian basis sets can be used to describe the diffuseness of the positron wave function. \cite{swann_calculations_2018,hofierka_many-body_2022,hofierka_gaussian-basis_2023,cassidy2024many,arthur2024positron}  Finally, placing ghost atoms around the ion or molecule allows to accurately account for virtual-Ps formation. \cite{hofierka_many-body_2022,hofierka_many-body_2022,hofierka_gaussian-basis_2023,cassidy2024many,arthur2024positron}
While this approach significantly improves the accuracy of the predicted binding, the number of required orbitals increases steeply. For example, in Ref.\citenum{hofierka_gaussian-basis_2023} the authors used 20 ghost atoms and over 1300 orbitals for H$^{-}$ and F$^{-}$. Moreover, the choice of the ghost atoms positions and of the even tempered basis set parameters increases the complexity of the calculation. We performed our POS-CCSD atomic binding energy calculations using the basis set from Ref.\citenum{hofierka_gaussian-basis_2023} to compare to the other results in the literature. All information needed to validate the findings in this paper are reported in the Zenodo online repository.\cite{riso2025capturing} The computational cost of a POS-CCSD calculation scales as the seventh power of the number of orbitals ($N^{7}$). Therefore, the calculation of binding energies even for these small systems is demanding and in some cases unfeasible. To tackle this problem, we extended our POS-CCSD methodology to include an active space framework. This allows us to restrict the number of electronic excitations included in the $T_{1}$ and $T_{2}$ operators, as well as the number of electron-positron excitations included in $S_{1}$, $S_{2}$ and $\Gamma$. In Tab. \ref{tab:Ghost_atoms} we report the values of the POS-CCSD binding energies for different active space dimensions (both for H$^{-}$ and F$^{-}$). The active space dimension N denotes that the first N canonical orbitals of the electron and positron that have been used in the cluster definition (see the light blue squares in Fig.\ref{fig:CC_and_states}). 
While this choice significantly improve the predicted binding (which is exactly the same as the reference value for F$^{-}$ and much closer to the reference 7.11 eV for H$^{-}$), we also note that the active space dimension required to obtain accurate results is very high. This is in agreement with the findings of Ref. \citenum{hofierka_many-body_2022} where orbitals up to 150 eV need to be included in the calculation to converge the binding energy. 
For neutral polyatomic molecules, where the positron is expected to be even more delocalized, the number of ghost atoms and the active space dimension required to obtain accurate binding energies will likely be even larger. We point out that previous relativistic L-p-CC calculations on the positron binding for neutral atoms have been performed in Ref. \citenum{dzuba2012relativistic}. While such systems are not conceptually different from the ions described above, the number of Gaussian orbitals required to describe the highly diffuse nature of the positron wave function exceeds our current computational capabilities (i.e. 2300 orbitals required using the basis set from Ref.\citenum{hofierka_gaussian-basis_2023}). In Ref. \citenum{dzuba2012relativistic}, which focused on positron-atom bound states, the authors instead used a much more efficient single-centre B-spline basis, which enables the angular integration and summation to be done analytically.
Alternative approaches have also been developed for positronic systems. Explicitly correlated methods, including explicitly correlated Gaussian formulations, provide an efficient description of the electron–positron cusp and short-range correlation, significantly reducing the need for large orbital basis sets\cite{brorsen_calculation_2017}. In contrast, diffusion Monte Carlo (DMC) solves the Schr\"odinger equation in real space rather than within a basis expansion\cite{cassella2024neural,mella2001stability}. These methods offer complementary routes to positron binding, but introduce distinct computational bottlenecks and approximations\cite{klopper2006r12,austin2012quantum}.

\subsection{Binding energies in polyatomic molecules}
\begin{figure}[!ht]
    \centering
    \includegraphics[width=\linewidth]{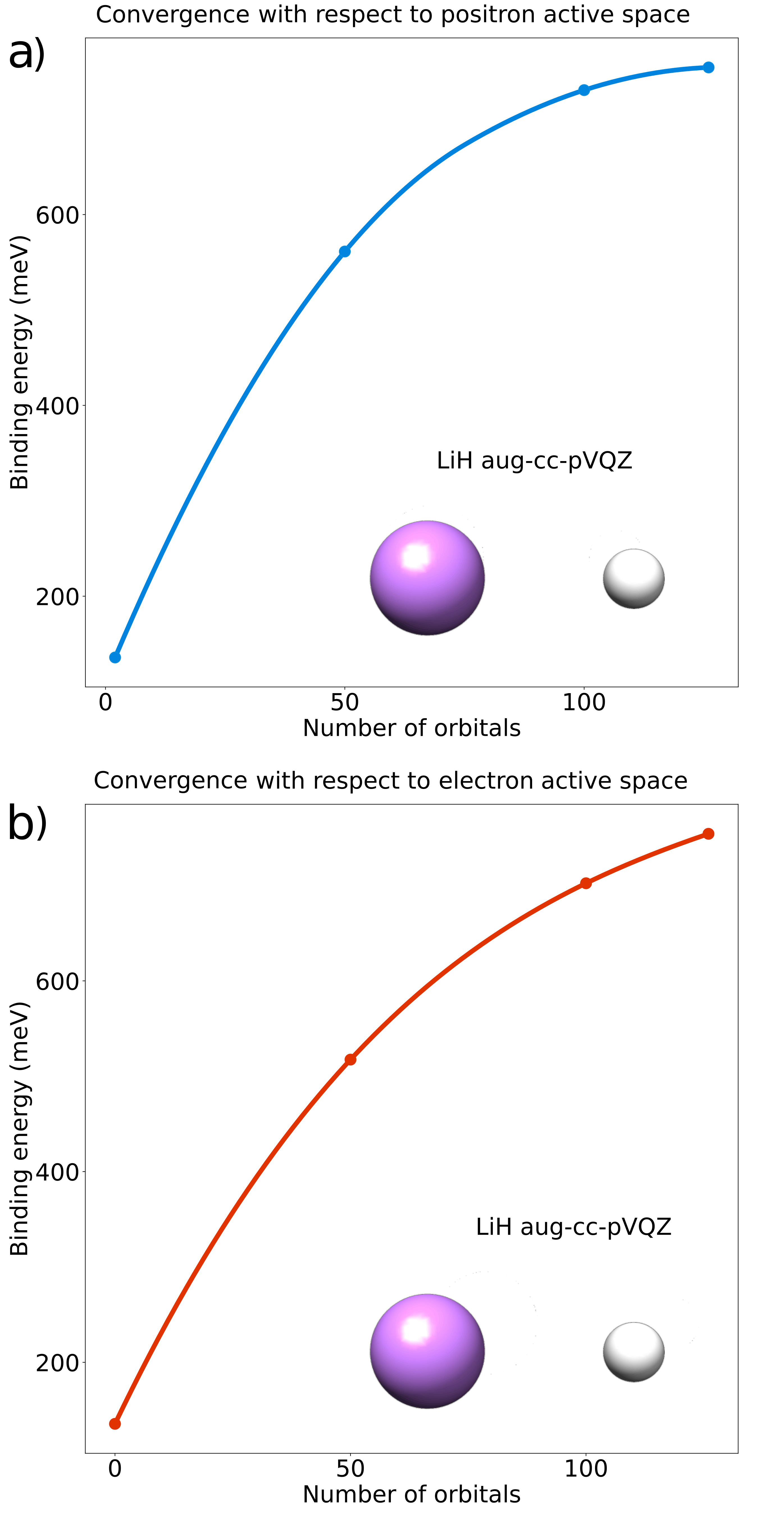}
    \caption{Difference in convergence to the full space result for the positron and electron active space selection for a LiH molecule at the aug-cc-pVQZ level. In Fig.\ref{fig:Active_space_in_QZ}a the full electronic active space is used in the calculation while in Fig.\ref{fig:Active_space_in_QZ}b the full positron active space is included.}
    \label{fig:Active_space_in_QZ}
\end{figure}

Positron binding to polyatomic molecules has been predicted and observed across different classes of systems including aromatic rings, alkanes, halogenated hydrocarbons, and small inorganic polar and non-polar compounds.\cite{danielson2010dipole,danielson2012interplay,gribakin2010positron,danielson2025improved,arthur2024positron,cassidy2024many,danielson2023positron} 
In Tab.\ref{tab:Polyatomic_merged}, we report the positron binding energies computed using different active space dimensions for a small selection of molecules across all classes. Moreover, in Tab. \ref{tab:Polyatomic_merged} we also report the the molecular dipole, polarizability and ionization energy. While insufficient to determine the binding energy, these are useful quantities to estimate how strongly a positron binds to a molecule.\cite{cassidy2024many} 
A dipole of >$.1.625$ D supports a positron bound state at the static level of theory \cite{Crawford67}. Thus, molecules with larger dipole moments can be expected to bind positrons, with correlation effects enhancing the binding energies. 
%The molecular dipole, indeed, correlates with binding at the static level; 
The isotropic polarizability is connected to the dynamical electron-positron correlation; the long-range positron-molecule potential is of the form $-\alpha_d/2r^4$, where $\alpha_d$ is the isotropic polarisability (assuming an isotropic system), and the ionization energy correlates with both the strength of the virtual-Ps formation process and how easy it is for the positron to perturb the electrons. Similarly to what discussed previously, the ghost atoms position and basis have been taken from Refs. \citenum{hofierka_many-body_2022,arthur2024positron}.
\begin{table*}[!ht]
\centering 
\caption{Dipole moment, isotropic polarizability, ionization energy, and positron binding energies for a set of polar and apolar polyatomic molecules. Molecular properties are computed at the CCSD level of theory using the geometries and basis sets from Ref.\cite{hofierka_many-body_2022}. Binding energies (in meV) are computed using POS-CCSD with different active space dimensions. Even in the larger active space calculations the POS-CCSD results are not converged, See Fig.~\ref{fig:dispersion}, and thus fail to reach the accuracy of $\Sigma^{GW+\Gamma+\Lambda^{\dagger}}$ (Refs.\citenum{hofierka_many-body_2022},\citenum{arthur2024positron}) or Configuration Interaction with Quantum Monte Carlo.\cite{upadhyay2024capturing} Significant increases of the predicted binding energies are observed when the T$_{2}$ operator is neglected. We highlight that the apparent improvement in the accuracy of the binding energies when T$_2$ is neglected is misleading, as the wave function is described less accurately. The sporadic agreement between the no-T$_{2}$ results and the experimental or theoretical references is to be regarded as accidental, in particular considering that the calculations omitting T$_2$ are also not converged and that including more orbitals in the active space of the calculation would be expected to change the results.}
\begin{tabular}{l c c c c c c c c}
\hline
\textrm{Molecule} & Dipole (D) & Polarizability (Å$^{3}$) & Ionization (eV) & Exp. & Theor. Ref. & 300 Orb. (no-T$_2$) & 300 Orb. & Larger Active Space \\
\hline
LiH & 5.886 & 4.275 & 7.985 & -- & 1060 [\citenum{hofierka_many-body_2022}] & 1011.6  & 825  & 909 \footnotemark[1] \\

Acetonitrile & 3.942 & 4.420 & 12.575 & 180 $\pm$ 10[\citenum{hofierka_many-body_2022}] & 207[\citenum{hofierka_many-body_2022}] & 243.8 & 130  & 155 \footnotemark[1] \\

Hydrogen cyanide & 3.021 & 2.426 & 13.922 & -- & 63--73 [\citenum{hofierka2024many}] & 78.6  & 28  & 37.8 \footnotemark[1] \\

Formaldehyde & 2.290 & 2.582 & 10.831 & -- & 28[\citenum{hofierka_many-body_2022}] & 28.6  & -3.1  & 5.6 \footnotemark[1] \\

Benzene & 0 & 10.437 & 9.135 & 132 $\pm$ 3[\citenum{arthur2024positron}] & 148[\citenum{arthur2024positron}] & 212  & 33  & 37 \footnotemark[2] \\

CS$_2$ & 0 & 9.021 & 10.059 & 75 $\pm$ 10[\citenum{danielson2010dipole}] & 87 $\pm$ 15[\citenum{upadhyay2024capturing}] & 144 & 11 & 24 \footnotemark[3] \\
\hline
\end{tabular}

\footnotetext[1]{500 orbitals in both electronic and positron space.}
\footnotetext[2]{300 orbitals for electrons and 400 orbitals for positron space.}
\footnotetext[3]{400 orbitals in both electronic and positron space.}

\label{tab:Polyatomic_merged}
\end{table*}
The maximum active space dimension used in our calculations are determined by our current implementation limitations ($\sim$2 TB of memory). 
The predicted binding energies are significantly different from either the experimental values or the theoretical reference. As a general trend, the $\varepsilon_{b}$ computed using POS-CCSD is lower than the theoretical reference values and it increases with the active space dimension, but overall the results are unconverged (see below). Moreover we note that, as expected, POS-CCSD performs best when the molecule has a more pronounced dipole, like for LiH and acetonitrile. For less polar systems (i.e. formaldehyde) smaller active spaces can give a negative binding energy (i.e., lack of binding). 
\begin{figure}[!ht]
    \centering
    \includegraphics[width=0.75\linewidth]{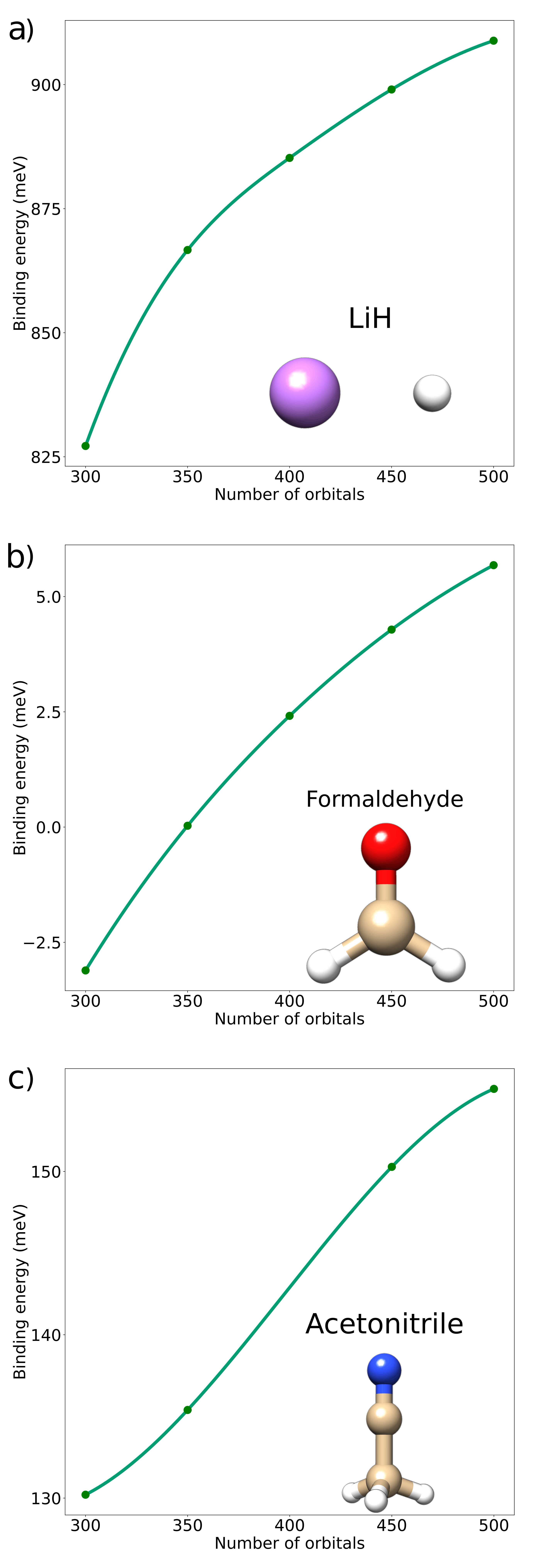}
    \caption{POS-CCSD calculated positron binding energies for LiH, formaldehyde and acetonitrile vs. number of orbitals included in the active space. Converged calculations require larger active spaces than are currently feasible with our current implementation and computational resources. Orbital energies of up to $\sim 150$ eV are included in the 500 orbital active space for all molecules.}
    \label{fig:dispersion}
\end{figure}
To check whether the POS-CCSD results are converged with respect to the dimension of the active space, in Figs.~\ref{fig:dispersion}a--c we plot the POS-CCSD calculated binding energy for LiH, acetonitrile and formaldehyde against the active space dimension, increasing the number of orbitals in both the electron and positron active space. Our results show that in all systems the binding curve has not reached a plateau, meaning that the full space binding is expected to be different even from the larger active space results reported in Tab.\ref{tab:Polyatomic_merged}. This is largely expected to explain the discrepancy between the POS-CCSD results and previous theoretical results. In conclusion, the results reported in the Tabs. \ref{tab:H_ion_single}-\ref{tab:Polyatomic_merged}  highlight the deficiency of the chosen basis set, incapable of balancing between the simultaneous diffuseness of the positron wave function at long range and its accumulation around negatively charged areas of the molecular system. Moreover, we highlight that the energy-based active space selection scheme discussed in this work seems to be sub-optimal for the selection of the electronic active space. This is illustrated in Fig.\ref{fig:Active_space_in_QZ}b, where we plot the dispersion of the POS-CCSD binding energy for LiH at the aug-cc-pVQZ basis set against the number of electron orbitals included in the active space while the full positron orbitals are included in the calculation. We notice that while the aug-cc-pVQZ basis set is clearly not enough to describe the spacial profile of the positron wave function, the positron active space selection scheme in Fig.\ref{fig:Active_space_in_QZ}a reaches convergence much faster than in Fig. \ref{fig:Active_space_in_QZ}b.  

To elucidate the relevance of electron correlation in the description of positron-molecule systems, we repeat the 300 active orbital calculations for the molecules listed in Tab.~\ref{tab:Polyatomic_merged} by excluding the T$_2$ cluster operator from the wave function. In this approximation, the reference electronic energy is at the Hartree–Fock level, and the POS-CCSD wave function only includes the T$_1$ operator, which describes electronic polarization induced by the positron, the $\Gamma$ operator, which describes positron polarization due to the electrons, and the S$_{1}$ and S$_{2}$ operators which account for the dynamical electron-positron correlation. 

\noindent This approximation leads to pronounced changes in the predicted binding energies. For all the systems the binding energies are artificially increased and appear closer to the experiments or theoretical values than the full POS-CCSD results. This is, however, misleading, as it arises from an incomplete physical description rather than a more accurate modeling of the wave function. Moreover, we stress that the apparent agreement between the no-T$_{2}$ calculations and the theoretical values is accidental, in particular because the no-T$_{2}$ results are equally not converged with the active space dimension. Nevertheless, these findings suggest an essential role of electron correlation in positron binding and highlight the need for further development of \textit{ab initio} methods that treat electrons and positrons on an equal footing. 

In order to compare POS-CCSD with previous methodologies without the active space dependency, in Tab.\ref{tab:Qz_full} we report POS-CCSD binding energies together with $\Sigma^{GW+\Gamma+\Lambda^{\dagger}}$ in the full aug-cc-pVQZ basis set space for LiH, H$^{-}$ and acetonitrile. We stress that this basis is not large enough to achieve a converged binding energy in either theory. 
We notice that POS-CCSD always predicts a lower binding energy than both $\Sigma^{GW+\Gamma+\Lambda^{\dagger}}$ and the no-T$_{2}$ calculations (see Fig.\ref{fig:Active_space_in_QZ} for the convergence with respect to orbitals in LiH). The discrepancy between $\Sigma^{GW+\Gamma+\Lambda^{\dagger}}$ and POS-CCSD is unclear and could arise from two main factors. First, this disagreement could be due to lack of higher order excitations in the POS-CCSD electron and electron-positron space (with the exception for H$^{-}$), leading to an inaccurate description of the wave function. This could potentially be resolved by including higher excitation operators in the cluster expansion, for example T$_{3}$ and S$_{3}$. 
Another possibility is that the neglect of electron correlation in the target electronic structure in the many-body theory approach, i.e., the use of frozen-target Hartree-Fock orbitals in the $\Sigma^{GW+\Gamma+\Lambda^{\dagger}}$ self energy diagrams, leads to an overestimation of binding in this basis. Self-consistent many-body calculations in which dressed \emph{electron} propagators 
(i.e., calculated at $GW$@BSE level for the electron) are used to construct the positron-molecule self energy diagrams are in principle feasible and will be the subject of future work to assess their importance. 
\begin{table}[!ht]
    \centering
     \caption{Full space comparison between $\Sigma^{GW+\Gamma+\Lambda^{\dagger}}$, POS-CCSD and the no-T$_{2}$ binding energies in the aug-cc-pVQZ basis set. We note that in all the systems significant variance is observed between the methods.}
    \begin{tabular}{c c c c}
    \hline 
     System & $\Sigma^{GW+\Gamma+\Lambda^{\dagger}}$ & no-T$_{2}$ & POS-CCSD \\
    \hline
    H$^{-}$ & 7.39\;\textrm{eV}& 6.87\;\textrm{eV} & 6.49\;\textrm{eV} \\
    LiH & 926\;\textrm{meV} & 932\;\textrm{meV} & 754\;\textrm{meV}\\
    Acetonitrile & 19\;\textrm{meV} & 166\;\textrm{meV} & -66\;\textrm{meV}\\ 
    \hline 
\end{tabular}
\label{tab:Qz_full}
\end{table}
\subsection{Capture aided vibrational effects}
\begin{figure}
    \centering
    \includegraphics[width=0.5\textwidth]{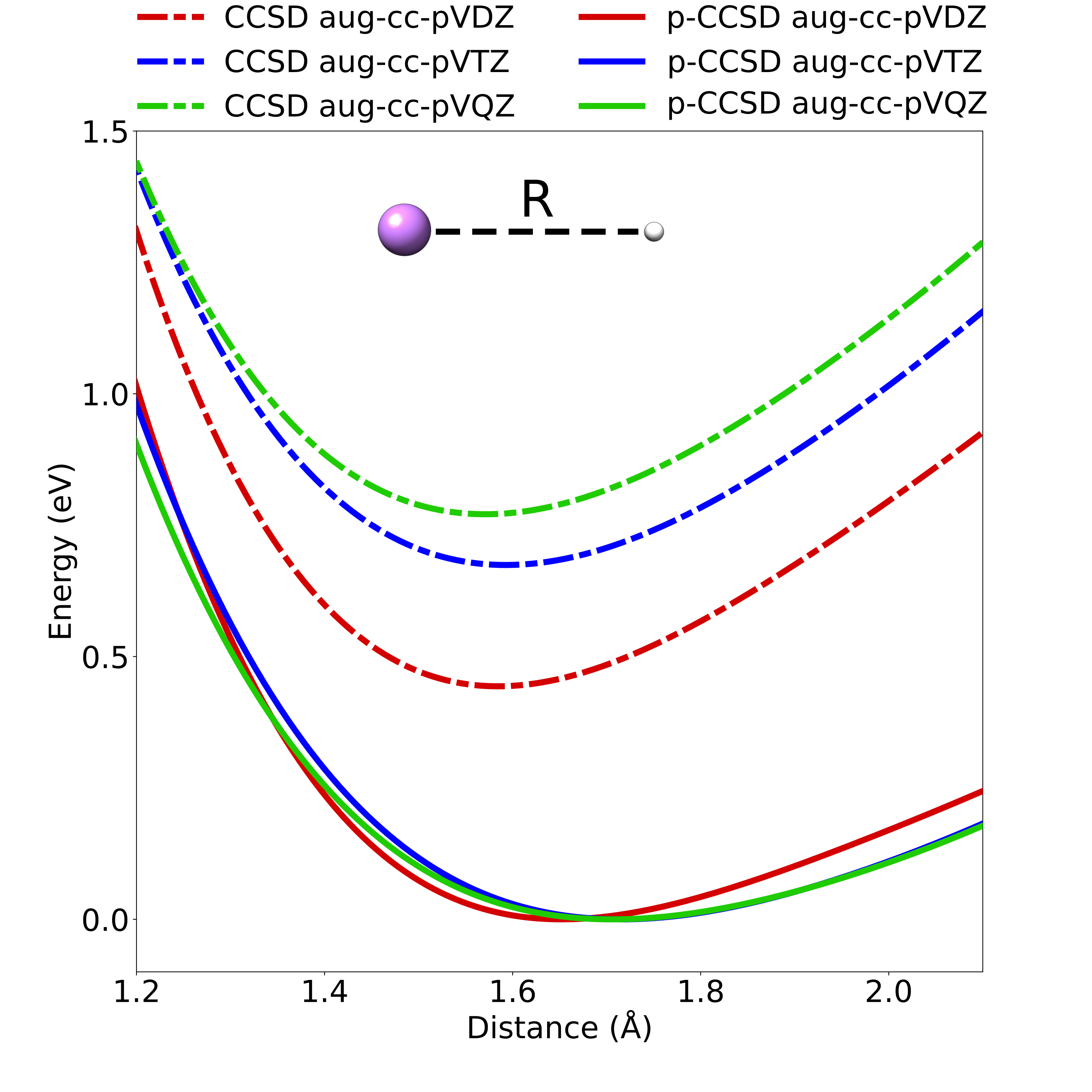}
    \caption{PES for LiH with and without positron attachment. We notice that the presence of the positron moves the equilibrium minimum to larger distances for every basis set. Each surface is shifted by the energy of the minimum of the POS-CCSD calculation for a given basis.}
    \label{fig:PES_LiH}
\end{figure}

The strong electron-positron interaction can modify the nuclear potential energy surface (PES) of the positron-molecule system relative to that of the bare molecule, see e.g.,\cite{gianturco2006positron,koyanagi2013positron,Kita2014,takayanagi2017quantum}.
In some cases, positron attachment might thus feasibly initiate chemical reactions because the nano/micro second lifetime of the meta-stable positron state is long enough time for nuclear reconfiguration before annihilation.
In Fig.\ref{fig:PES_LiH}, we study the modification of the PES in LiH after positron attachment using different standard basis sets. 
For each basis set, the POS-CCSD and CCSD curves have been shifted by the energy of the POS-CCSD minimum, a result already observed in  Refs.\cite{gianturco2006positron,iida2021contribution}. 

\begin{figure*}
    \centering
    \includegraphics[width=0.75\textwidth]{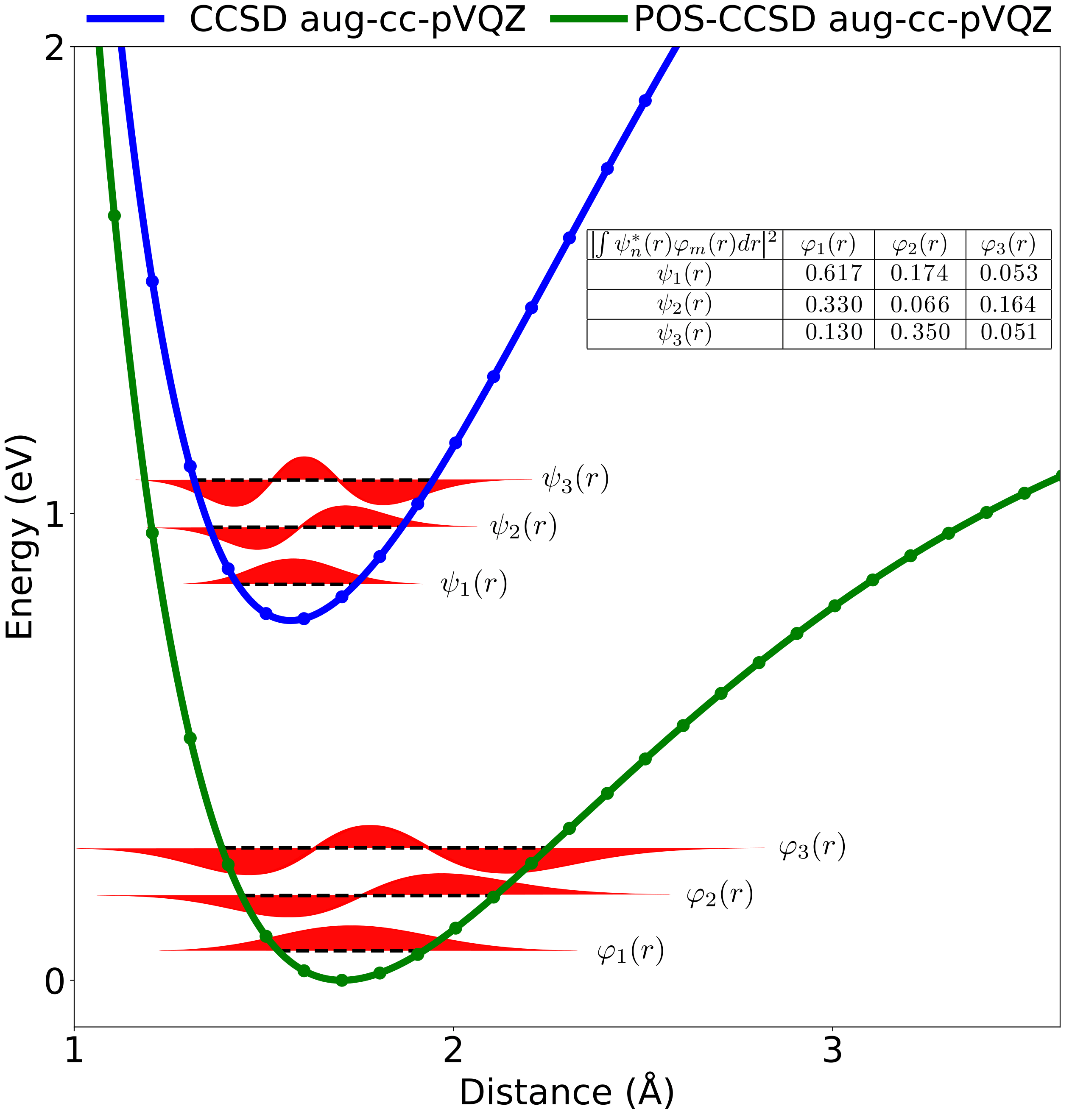}
    \caption{Vibrational states of the POS-CCSD and CCSD aug-cc-pVQZ PES. The overlap matrix between the vibrational states is reported in the top right table (inset).}
    \label{fig:Overlap}
\end{figure*}

Regardless the choice of basis set, the PES is significantly modified by the presence of the positron.
The vibrational structure of the molecule also changes. The vibrational energies in cm$^{-1}$ are reported in Tab.\ref{tab:Vib_energy}. They were computed using the VIBROT module of the OpenMolcas program\cite{aquilante2020modern}. For all considered basis sets, the presence of the positron reduces the frequency of the vibrational states. This is visually confirmed by the fact that the PES becomes flatter around the equilibrium geometry. Differences in the spacing of the levels are clustered around 80 cm$^{-1}\approx$ 10 meV, which is a sizeable quantity compared to $\varepsilon_{b}$. In Fig.\ref{fig:Overlap} we show the aug-cc-pVQZ PES. Once the positron has been captured, the multicomposite system will evolve following the green POS-CCSD curve. The dynamics depends on the initial vibrational state of the molecule and on the overlap elements between the vibrational states of the CCSD and the POS-CCSD curves. For example, using the numbers in Fig.\ref{fig:Overlap}, a lithium hydride in its vibrational ground state ($\psi_{1}$) is more likely to go into the ground state of the positron-molecule complex ($\varphi_{1}$), while if the molecule is in its second excited state ($\psi_{3}$) the system will evolve most likely as the first excited state of the green curve ($\varphi_{2}$). Such effects have been studied for the case of positron capture in HCN in Ref. \citenum{takayanagi2017quantum}. Since the lifetime of the positron-molecule complex is long compared to the nuclear motion, the positron induced relaxation effects, Fig.\ref{fig:Overlap}, might be used to initiate ground state reactivity. Moreover, we point out that since the positron is mostly localized around the negatively charged areas of the molecule, this effect, in addition to the resulting localised annihilation\cite{koyanagi:2013,hofierka_many-body_2022}, has the potential to selectively activate desired areas of a polyatomic system. 

\begin{table*}[!ht]
\centering
\caption{Vibrational energies in cm$^{-1}$ for the LiH PES with and without the positron capture. We notice that while the levels are not fully converged with the basis set, the vibrational energies for the positron case are much lower than in the no-positron case.}
\begin{tabular}{l l l l l l l }
\hline 
Level & \multicolumn{3}{c }{No-positron} & \multicolumn{3}{c }{Positron} \\
\hline
&    aug-cc-pVDZ & aug-cc-pVTZ & aug-cc-pVQZ & aug-cc-pVDZ & aug-cc-pVTZ & aug-cc-pVQZ \\ 
\hline
0 &  643.719\;cm$^{-1}$ &  627.477 \;cm$^{-1}$&  591.264 \;cm$^{-1}$ &  583.583 \;cm$^{-1}$&  529.372 \;cm$^{-1}$&  510.278 \;cm$^{-1}$ \\
1 & 1756.016 \;cm$^{-1}$& 1731.435 \;cm$^{-1}$& 1579.778 \;cm$^{-1}$& 1639.036 \;cm$^{-1}$& 1518.106 \;cm$^{-1}$& 1463.676 \;cm$^{-1}$ \\
2 & 2662.007 \;cm$^{-1}$& 2643.352 \;cm$^{-1}$& 2394.468 \;cm$^{-1}$& 2528.786 \;cm$^{-1}$& 2371.423 \;cm$^{-1}$& 2286.925 \;cm$^{-1}$ \\
\hline 
\end{tabular}
\label{tab:Vib_energy}
\end{table*}
\subsection{Challenges in comparing theoretical and experimental positron binding energies}
At present theoretical values of the binding energy are computed by performing single-point electron-positron calculations, that is at a fixed nuclear configuration. The binding energy can therefore be either computed by taking the energy difference 
\begin{equation}
\varepsilon_{b}=E_{\textrm{total}}(\textrm{molecule+positron})-E_{\textrm{total}}(\textrm{molecule})
\end{equation}
or by performing a positron quasi-particle calculation.\cite{hofierka_many-body_2022} In light of the results of the previous section, however, we point out that comparing single point results with experimental values entails disregarding the nuclear relaxation effects that are present in the experimental data. Previous theoretical efforts seem to indicate strong relaxation effects for dipolar systems ($\sim10$\% of the binding in Ref.\cite{gianturco2006positron,romero2014calculation} for alkali hydrides) and a small effect for oxides\cite{Buenker:2007,Buenker:2008} and organic compounds (few percentages of the binding in Ref.\cite{tachikawa2014positron} for a single mode C=O, C-N relaxation). A remaining question is whether nuclear relaxation effects become more significant in larger molecular systems, where a greater number of vibrational degrees of freedom may be influenced by positron capture. We point out, however, that no approach has been proposed yet to account for such effects and that the \textit{ab initio} calculation of both the Frank-Condon factors and the vibrational energy restructuring together with the electronic and positronic wave function is what is needed to provide an accurate modeling of positron attachment. These aspects will be the focus of future studies. 
We note that, however, that there are proposals\cite{surko2012measuring,swann2016formation} that circumvent the vibrational Feshbach resonant attachment process, and would enable observation of binding in the vibrational ground state, minimizing vibrational effects.
\section{Conclusion}
In this work, we presented a coupled cluster formalism to calculate positron binding energies in
polyatomic molecules. Contrary to other available methodologies, our POS-CCSD approach accounts for electron-electron and electron-positron correlation on the same footing, including electron polaritization through the $T_{1}$ cluster operator, electron-electron correlation through the $T_{2}$ cluster and electron-positron correlation through the $S_{1}$ and $S_{2}$ operators. While the accuracy of the proposed approach is quite high in the case of atomic ions, we notice that even for these small systems, a very large basis and additional ghost atoms were needed to reach a satisfactory binding energy. Moreover, the position of the additional ghost atoms and their basis set introduce additional complications in the calculations. These observations underline the urgent need for a consistent and systematic development of positron optimized basis sets. For polyatomic molecules, where even more orbitals and ghost atoms are needed to accurately model binding energies\cite{hofierka_many-body_2022}, the POS-CCSD results are not always close to the experimental values, with errors up to 80$\%$ of the experimental data. This is most likely due to an insufficient number of orbitals being included in the active space calculation. Nevertheless, we notice that a significant improvement is observed when the active space dimension is increased. Future efforts will therefore focus on the optimization of the memory requirements for the current POS-CCSD implementation. 

We also considered the effects of vibrations and nuclear relaxation for LiH. We found that the presence of the positron leads to a non-neglible modification of the PES, leading to large nuclear relaxation effects as previously discussed by Gianturco\cite{gianturco2006positron}, and by Tachikawa and colleagues for other molecules \cite{koyanagi2013positron,Kita2014,takayanagi2017quantum} The results suggest that caution should be exercised when comparing fixed-nuclei calculations to experiment, 
and more importantly that nuclear relaxation effects could be used to activate reactive groups or modify the dynamics of molecular excited states.
\subsection*{Supplementary Material}
In the supplementary materials we present the POS-CCSD equations and discuss the comparison between POS-CCSD and other methods for the calculation of positron binding energies.
\subsection*{Acknowledgements}
R.R.R, J.H.M.T, F.R. and H.K. acknowledge funding from the European Research Council (ERC) under the European Union’s Horizon 2020 Research and Innovation Programme (grant agreement No.~101020016). D.G.G. thanks G. F. Gribakin and A. R. Swann for useful conversations, and acknowledges funding from the European Research Council (ERC) under the European Union’s Horizon 2020 Research and Innovation Programme (grant agreement No.~101170577).
\subsection{Data availability statement}
The input and output files that support the findings in this work can be found in the Zenodo repository \cite{riso2025capturing}. The eT code is available upon request to the corresponding authors.
\bibliography{main}

\clearpage

\begin{figure}
    \centering
    \includegraphics[page=1,width=\textwidth]{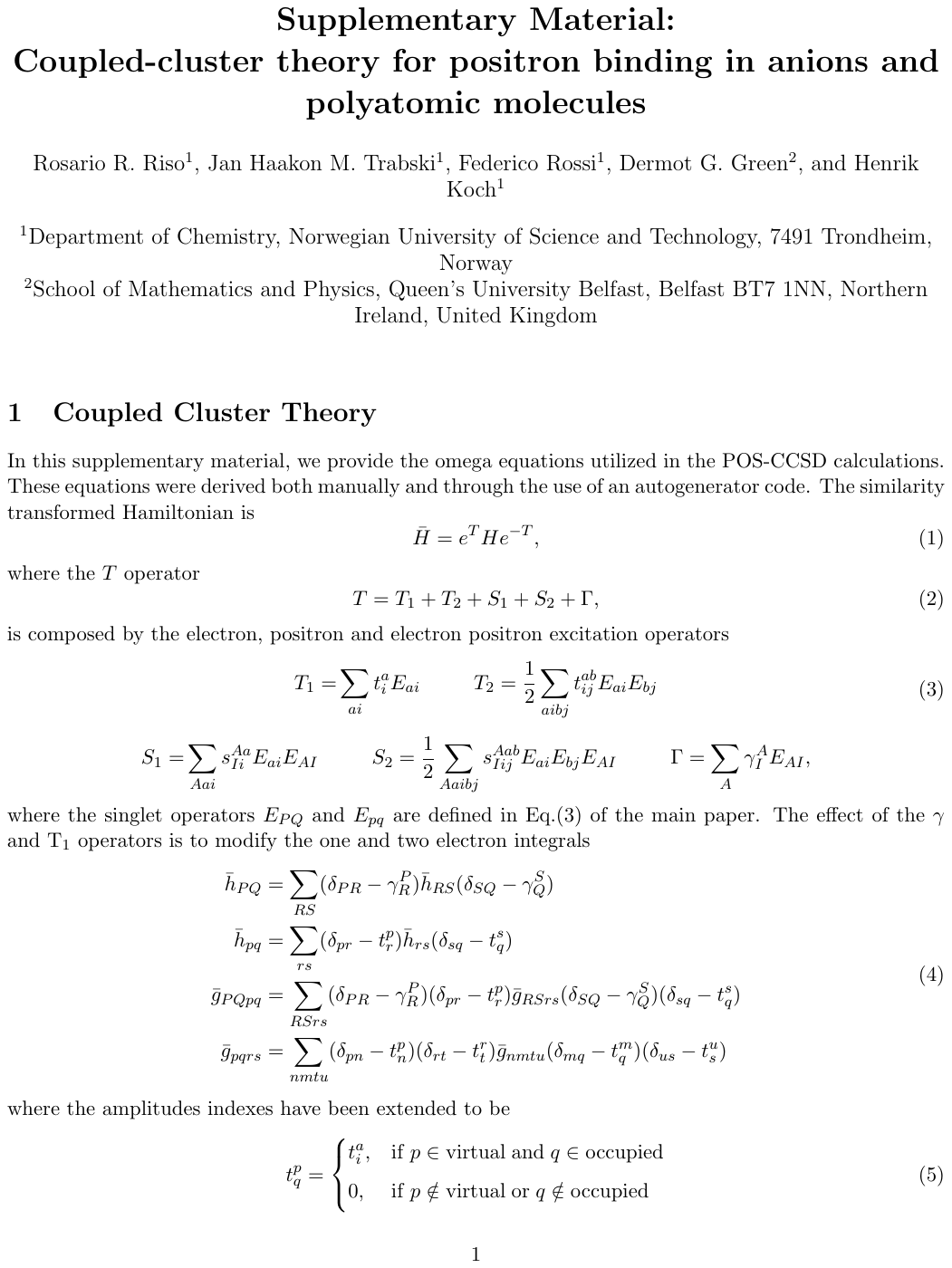}
\end{figure}
\begin{figure}
    \centering
    \includegraphics[page=2,width=\textwidth]{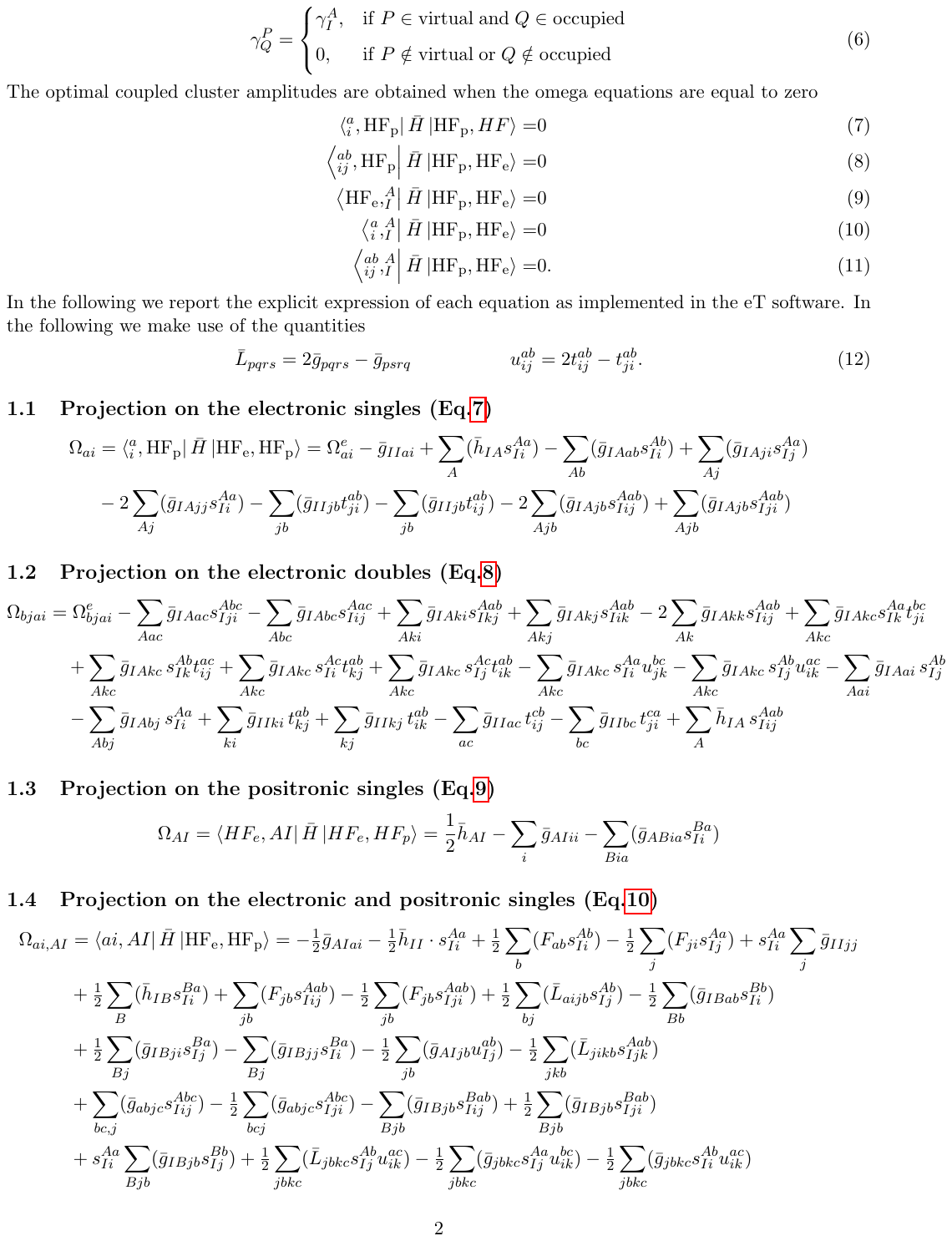}
\end{figure}
\begin{figure}
    \centering
    \includegraphics[page=3,width=\textwidth]{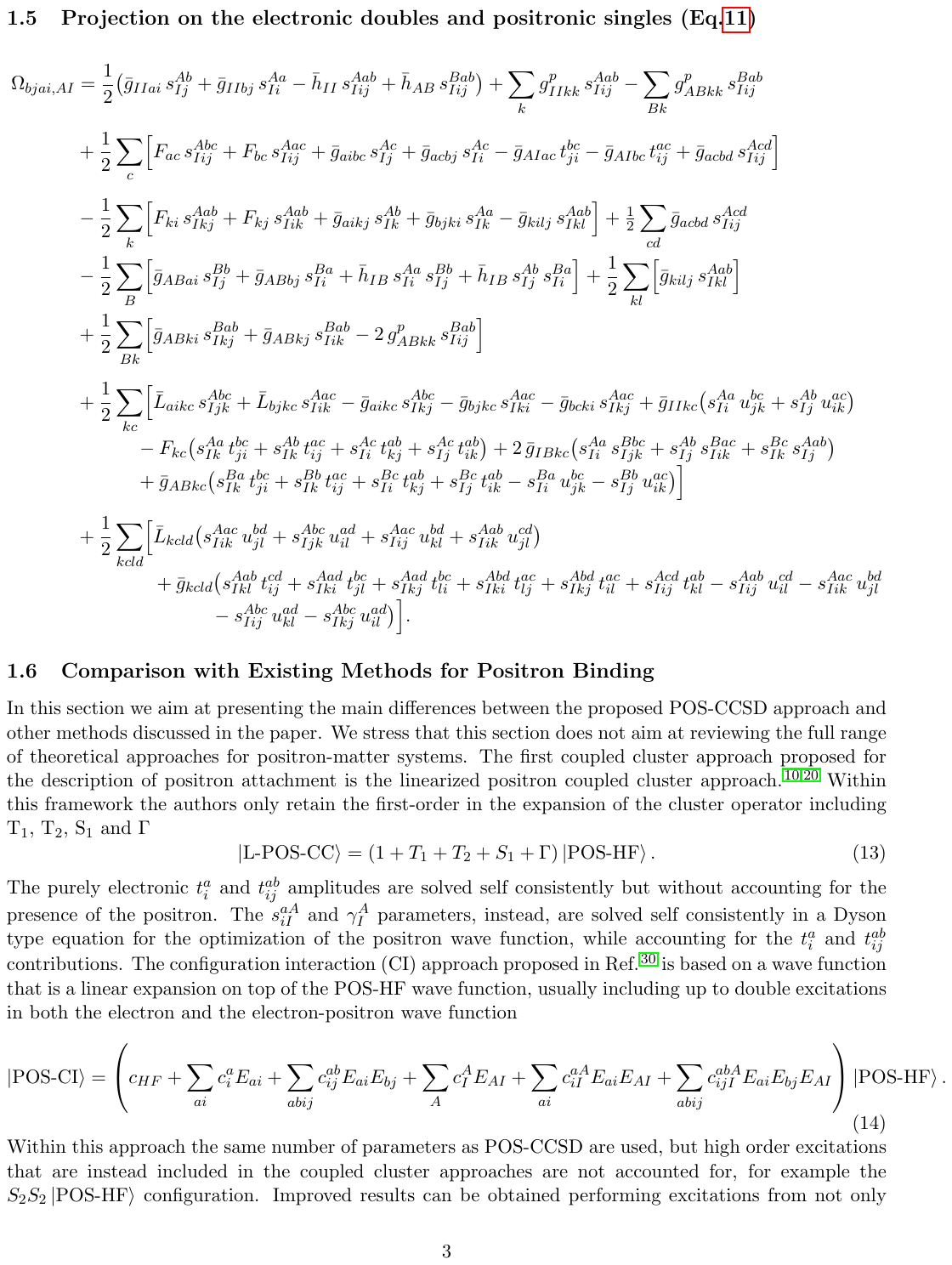}
\end{figure}
\begin{figure}
    \centering
    \includegraphics[page=4,width=\textwidth]{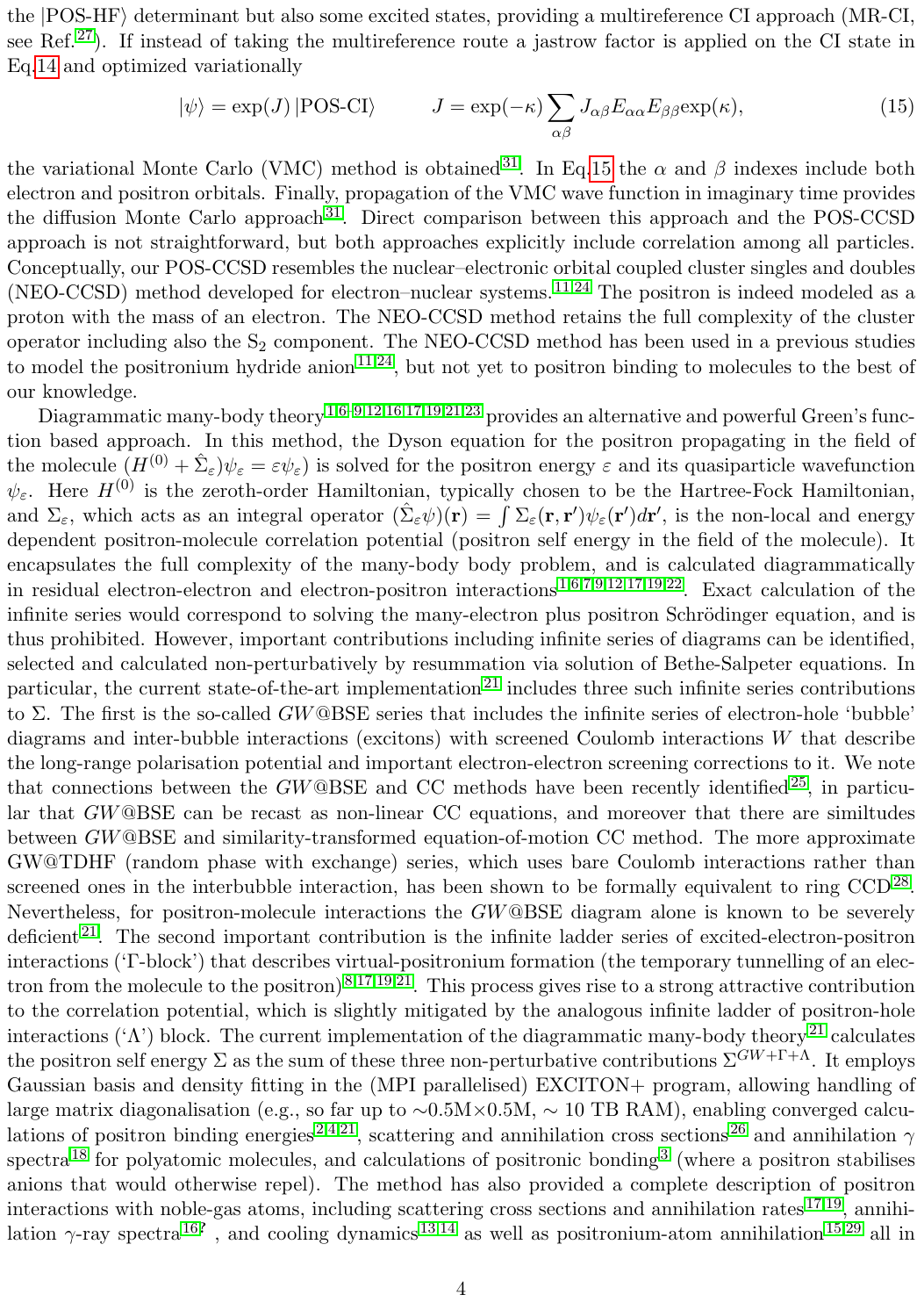}
\end{figure}
\begin{figure}
    \centering
    \includegraphics[page=5,width=\textwidth]{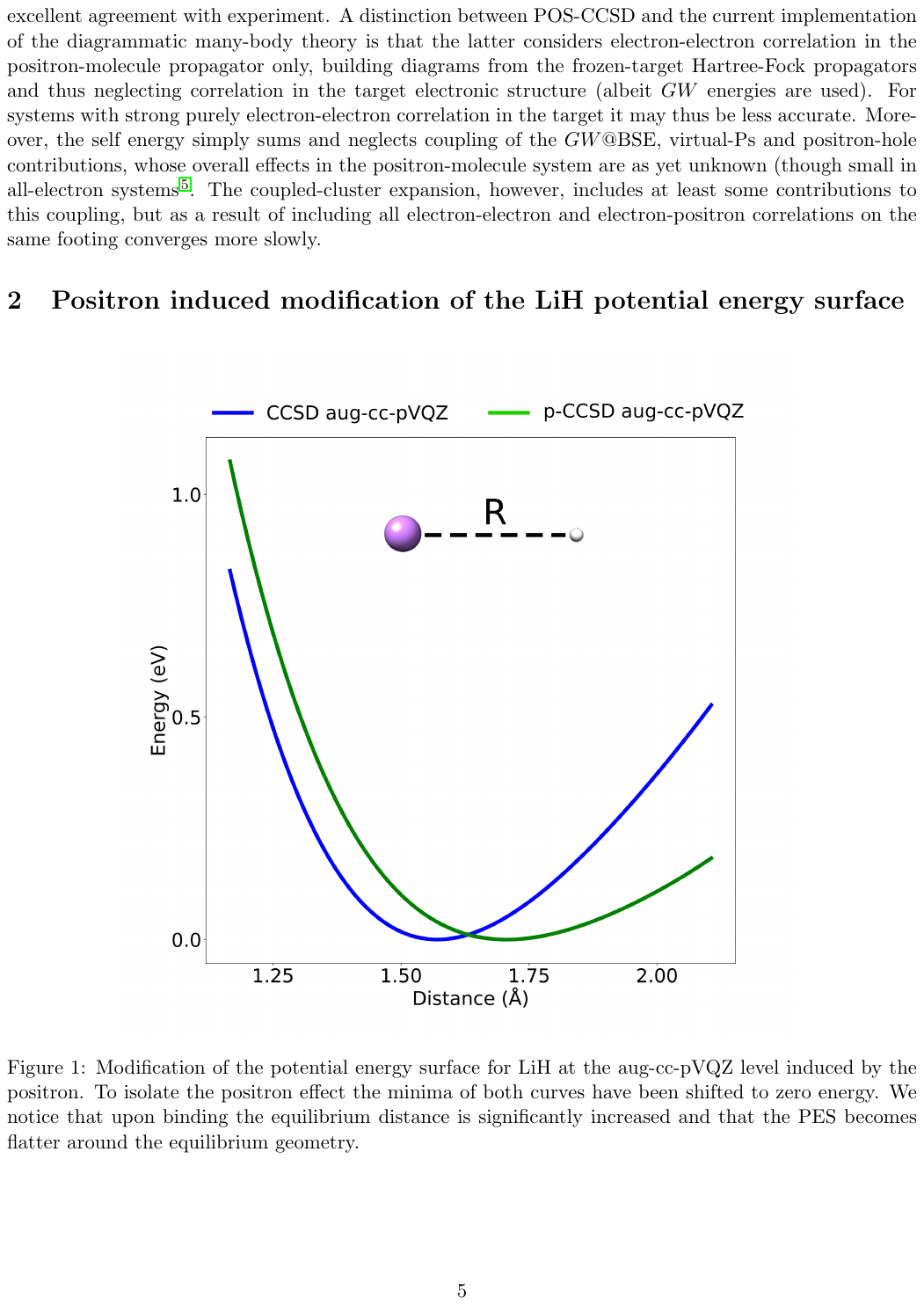}
\end{figure}
\begin{figure}
    \centering
    \includegraphics[page=6,width=\textwidth]{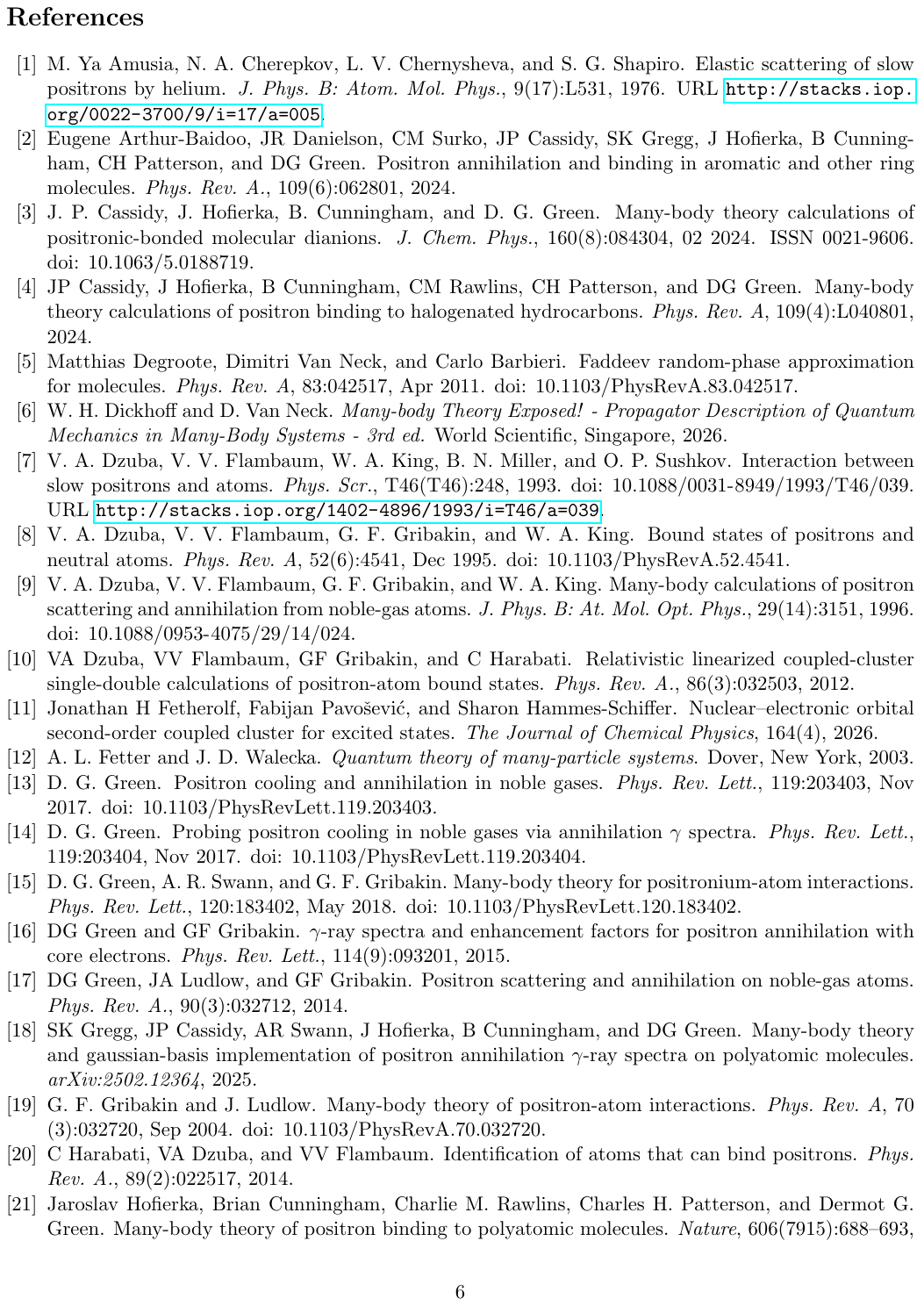}
\end{figure}

\end{document}